\documentclass[traditabstract,longauth,letter]{aa}
\usepackage{amsmath}
\usepackage{epsfig}
\usepackage{url}
\usepackage{hyperref}
\usepackage{tikz}
\usepackage{pgfplots}
\usepackage{natbib}

\def\email#1{\href{mailto:#1}{\url{#1}}}

\newcommand{\diag}{\mathrm{diag}}

\newcommand{\degree}{$^\circ$}

\def\argmin{\mathop{\mathrm{arg\,min}}}
\def\diag{\mathop{\mathrm{diag}}}

\def\Argmin#1#2{\displaystyle \argmin_{#1}{#2}}

\input{dep/alphabet}
\graphicspath{ {./figures/} }

\title{Feasibility and performances of compressed-sensing and sparse
  map-making with Herschel/PACS data}

\author{Nicolas Barbey$^1$, Marc Sauvage$^1$, Jean-Luc Starck$^1$,
  Roland Ottensamer$^2$, Pierre Chanial$^1$}

\institute{$^1$SAp/Irfu/DSM/CEA, Centre d'études de Saclay, Orme des
  Merisiers, Bat 709, 91191 Gif sur Yvette, France \\ $^2$University
  of Vienna, Department of Astronomy, T\"urkenschanzstr. 17, A-1180
  Wien, Austria }

\authorrunning{Nicolas Barbey \textit{et al.}}

\titlerunning{Compressed-sensing and sparse map-making with
  Herschel/PACS}

\keywords{nstrumentation: photometers, Methods: numerical,Methods:
  statistical, Techniques: image processing, Galaxies: general}

\abstract{The Herschel Space Observatory of ESA was launched in May
  2009 and is in operation since.  From its distant orbit around L2 it
  needs to transmit a huge quantity of information through a very
  limited bandwidth. This is especially true for the PACS imaging
  camera which needs to compress its data far more than what can be
  achieved with lossless compression. This is currently solved by
  including lossy averaging and rounding steps on board. Recently, a
  new theory called compressed-sensing emerged from the statistics
  community. This theory makes use of the sparsity of natural (or
  astrophysical) images to optimize the acquisition scheme of the data
  needed to estimate those images. Thus, it can lead to high
  compression factors.
  
  A previous article by Bobin et al. (2008) showed how the new theory
  could be applied to simulated Herschel/PACS data to solve the
  compression requirement of the instrument. In this article, we show
  that compressed-sensing theory can indeed be successfully applied to
  actual Herschel/PACS data and give significant improvements over the
  standard pipeline. In order to fully use the redundancy present in
  the data, we perform full sky map estimation and decompression at
  the same time, which cannot be done in most other compression
  methods.  We also demonstrate that the various artifacts affecting
  the data (pink noise, glitches, whose behavior is a priori not well
  compatible with compressed-sensing) can be handled as well in this
  new framework. Finally, we make a comparison between the methods
  from the compressed-sensing scheme and data acquired with the
  standard compression scheme. We discuss improvements that can be
  made on ground for the creation of sky maps from the data.}

\begin{document}

\maketitle

\section{Introduction}
\label{sec:intro}

The Herschel Space Observatory \citep{pilbratt2010herschel} is a three
and a half year mission carrying instruments to observe in the
far-infrared. It is dedicated to the investigation of galaxy formation
and evolution mechanisms, star formation and interaction with the
interstellar medium, molecular chemistry in the universe and finally
chemical analysis of the atmospheres of bodies of the Solar
System. With its 3.5\,m telescope, Herschel is the largest space
observatory to date.  Its three instruments are: the Photodetector
Array Camera and Spectrometer (PACS)
\citep{poglitsch2010photodetector}, the Spectral and Photometric
Imaging Receiver (SPIRE), \citep{griffin2010spire}, and the Heterodyne
Instrument for the Far Infrared (HIFI) \citep{degraauw2010hifi}.

Due to power and dissipation constraints, routine operations on
Herschel see only one instrument powered per observing day (OD). A
special mode exists, called the SPIRE$+$PACS parallel mode (see
below), where both SPIRE and PACS can be operated. The duty-cycle of
Herschel rests on the concept of the OD, which typically consists in a
period of autonomous observations of 20-22\,h, and a daily
telecommunication period (DTCP) of 2-4\,h during which the data
recorded on board is downlinked and the program of the next ODs is
uplinked (Herschel always stores on-board the program of its next two
ODs, in case of problems during the DTCP).

Limits to the amount of data that can be generated by the instruments
are imposed by (1) the speed of the internal communication links
between the instruments and the service module, and (2) the speed and
duration of the communication link between the satellite and its
ground stations. These limits translates typically in an acceptable
average data rate for the instrument during the OD of 130\,kbit/s,
while the PACS photometer, featuring the largest number of pixels
($\sim$2500) and a high sampling frequency (40\,Hz) provides a native
data rate of 1600\,kbit/s, to which 40\,kbit/s have to be added for
housekeeping information. The situation is even more complex for the
PACS spectrometer at 4000\,kbit/s (but the integration ramps can be
compressed on-board in a natural fashion), while SPIRE and HIFI have
much smaller data rates around or below 100\,kbit/s, requiring no
compression on board.

We focus on the PACS imaging camera, which is made of two arrays of
bolometers operating in the 55 to 210 $\mu$m range. Due to the above
restrictions PACS data need to be compressed by a factor of 16 to
32. For this purpose, on-board data reduction and compression is
carried out by dedicated sub-units \citep{ottensamer2008spie}.

In this paper, we investigate alternative compression modes to attain
such a high compression factor without sacrificing noise or
resolution. The study in this paper is based on the theory of
compressed-sensing and represents a real-world implementation of ideas
presented in \citet{bobin2008compressed}. Simply put,
compressed-sensing is a promising theory which leverages the sparsity
properties of data to allow measurements beyond the Shannon-Nyquist
limit \citep{donoho2006compressed, candes2008restricted}. In the
context of PACS, it offers potential to optimize on-board compression.
It has the very strong advantage that it is not computationally
expensive on the acquisition side, a very important point for a space
mission. An additional constraint for PACS is that CS had to be
implemented after the conception of the instrument, while the
instrument hardware cannot be changed as well as most of the on-board
software.

We first present the PACS instrument, describe the properties of its
data, and explain how we model the instrument. Then, we explain how we
perform map-making using PACS data which is a central stage in data
analysis.  Finally, we introduce our compressed-sensing implementation
and we show how the map-making process can be modified in the
compressed-sensing framework to estimate state-of-the-art or better
sky maps.

\section{PACS data}
\label{sec:data}

\subsection{The PACS imaging camera}
\label{subsec:pacs}

The PACS imaging camera is made of two arrays of bolometers. One array
is made of two $16 \times 16$ matrices forming a $32 \times 16$
array. It is associated with the ``red'' filter ranging from 130 to
210 $\mu$m. The other array is made of eight $16 \times 16$ matrices
and thus has a shape of $64 \times 32$ pixels. It is associated with
the filters ``green'' and ``blue'' ranging respectively from 85 to 130
$\mu$m and from 60 to 85 $\mu$m. Figure (\ref{fig:data}) shows a
single frame of the PACS blue array.

The PACS camera can be used in two modes: the chop-nod mode and the
scan mode. The chop-nod mode consists in using an internal chopper to
alternate between two sky positions (ON and OFF) and the satellite to
exchange the ON and OFF beam positions. It was originally dedicated to
the observation of point sources, with the chop-nod modulation used to
get rid of the low-frequency noise originating in the
detector. However it turned out that even for point sources, scan-mode
observations are more efficient in terms of achieved sensitivity for a
given observing time. Therefore this mode is not considered here. The
scan mode scans the sky in order to observe larger areas than the
detector field-of-view. When PACS is the only instrument powered, the
prime mode, there are three possible scan speeds: 10, 20 and 60
arcsec/s. The PACS camera can also work in scan mode in conjunction
with SPIRE, the parallel mode, in which case observations are driven
by SPIRE requirements and the scan speed is either 30 or 60
arcsec/s. Most of the PACS data are therefore obtained in scan mode,
either in prime or parallel mode, and in any case, the bolometer
arrays are read at 40\,Hz.

The data are compressed using a lossless entropic compression scheme
which yields up to a factor of 4, requiring another factor of 4 (8 in
parallel mode) to be achieved with lossy steps. For this purpose, 4
(or 8) consecutive frames are averaged before lossless
compression. Additionally, a bit-rounding step is performed. It
consists in rounding the averaging that is performed on board so that
the number of bits required to encode the data is reduced by 1 or 2
bits. This rounding is randomly up and down so as to not bias the
results. In prime mode, one bit is removed while in parallel mode two
bits are removed. We will not consider the bit-rounding operation in
this article. In parallel mode an averaging of 8 frames is required
because the satellite bandwidth is shared with SPIRE. In fast parallel
mode, this averaging results in a blurring of the frames and thus a
loss of resolution in the sky maps which is highly undesirable.
Indeed, in this mode the scan speed can be 60 arcsec/s, or 18.75
pixels/s. Frames are taken at 40 Hz, thus the displacement between two
frames is approximately 0.5 pixel. In parallel mode 8 frames are
averaged with a total motion of around 4 pixels. This is very
significant compared to the point-spread function (PSF) in the blue
filter which has a full-width at half maximum (FWHM) of 1.6 pixels.

This blurring is a very strong incentive to investigate other
compression schemes. Presently, it can be avoided only by configuring
the so-called transparent compression mode, where only one-eighth of
the frame area is transmitted, but at full temporal resolution.  With
this mode we can experiment with all the stages of the acquisition
while incorporating all features and artifacts of the actual
instrument. 

Indeed, care must be taken when using a new acquisition or compression
mode that it will not decrease our ability to separate the signal of
interest from instrumental artifacts. PACS imaging data have some
properties that have to be taken into consideration.  The most salient
one is a large signal offset that varies for each pixel.  The
dispersion of the offset among the detectors occupies half of the
range of values allowed by the 16 bits digitisation, whereas the
typical signal range is around 15 bits with a noise entropy of 4-6
bits. This noise measurement is not a theoretical value, but the
entropy of decorrelated measured data. It has been determined in two
ways : computing the Shannon entropy of the difference signal minus
0.5 bit (Differentiation increases the sigma by a factor $\sqrt(2)$ or
half a bit in entropy, which needs to be subtracted to get the
original value) and computing the Shannon entropy of the
high-frequency coefficients of a wavelet (for example Haar) transform.
Both methods give almost the same value. This is essentially the
readout noise, so it does not take into account the pink component of
the noise which can be seen at longer time scales. Hence, the
quantization noise is not negligible with respect to the signal and
adds to the other sources of noise. As can be seen in Figure
\ref{fig:data}a, the signal of interest is not visible when the offset
is not removed and we cannot subtract it on board. If the offset is
removed (Figure \ref{fig:data}b), it is still hard to detect physical
structure because of the noise. Some glitches can be identified
however that could not be seen in Figure \ref{fig:data}a. It
corresponds to the high-valued isolated pixels.

\begin{figure}
  \centering
  \includegraphics[width=\linewidth]{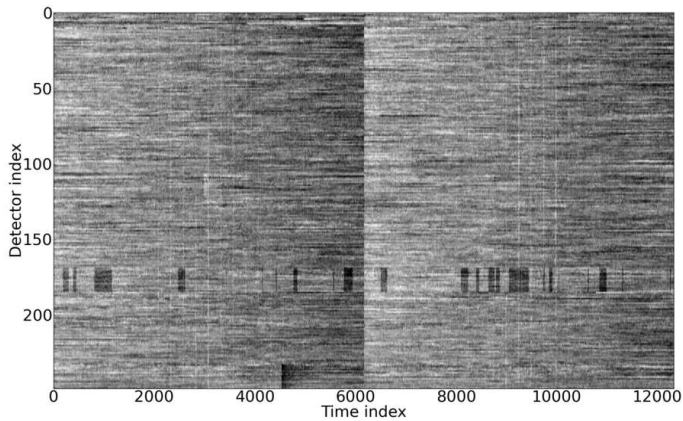}
  \caption{Data set of the PACS imaging camera in the blue band in the
    transparent mode. This set consists of two cross-scans. The offset
    has been removed. All pixels have been reordered in a vector shape
    and displayed as a function of time (so that each pixel is shown
    as a horizontal line). Horizontal structures are the drift of the
    offset. The discontinuity in the offsets at the center of the
    figure comes from the temporal gap between the two cross-scans.
    An unstable line (pixels 176 to 192) results in frequent intensity
    discontinuities along both scans. At time index 45000 the
    additional drift is due to a cosmic ray hit in the multiplexing
    electronics affecting a whole line. Finally, the faint vertical
    structures are due to the bright galaxy nucleus passing in front
    of the detectors.}
  \label{fig:tod}
\end{figure}

Furthermore, there is a drift of this offset (Figure \ref{fig:tod})
which can be seen as pink noise ($1/f$ noise). More precisely, it is a
$1/f^{1/2}$ noise as its power spectrum decreases with the inverse of
the square root of the frequency. One of the main issues in PACS data
processing is the separation of the signal from this drift. Its
amplitude is of the order of magnitude of the signal of interest. In
addition, this pink noise is affecting all the temporal frequencies in
the PACS bandwidth. In other words, the transition between pink noise
and white noise is happening at higher frequencies than the sampling
frequency of PACS.

Once the offset is removed, other features become visible, as can be
seen in Figure \ref{fig:tod}.  For instance, the bad behavior of a
whole line of pixels results in frequent discontinuities of the
intensity as a function of time (pixels 176 to 192 in the figure). A
cosmic ray hit in the electronics of the last line around the time
index 45000 results in an additional drift of these pixels. Other
cosmic rays can hit individual detectors and be the cause of wrong
intensity values (outliers). However, those cosmic ray hits are
affecting less than 1 percent of the data due to the spatial structure
of the detector elements (they are hollow). Outliers due to cosmic
rays are usually filtered out by comparing the set of data values
corresponding to one map pixel and masking values beyond a threshold
(see section \ref{subsec:deglitching} for details).

The faint vertical structures in Figure \ref{fig:tod} are in fact the
nucleus of the galaxy which passes in front of the detectors multiple
times. Only the nucleus is bright enough to be visible in the offset
removed data. Projection of the data onto the sphere of the sky as
well as filtering of the $1 / f^{1/2}$ noise are required in order to
see the fainter parts. After the projection, the contributions of each
pixel at each time index add up in order to greatly increase the
visibility of the fainter objects.

\subsection{Instrument model}
\label{subsec:model}

Our optimal reconstruction scheme requires an understanding of the
acquisition process, done here with an instrument model.

For the rest of this article, the data sets and the sky maps are
vectorized (i.e. re-indexed). Vectors are represented as bold
lower-case letters.  Linear operators are represented as matrices and
written in bold upper-case letters.

\begin{figure}
  \centering
  \input{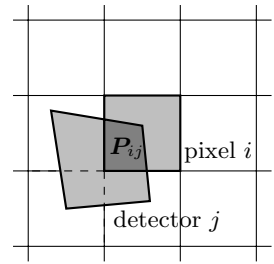}
  \caption{Sketch of the intersection of the sky map pixel $i$ and
    detector $j$. The matrix element $(i, j)$ of the projection matrix
    $\Pb$ is equal to the intersecting area of the pixel $i$ with the
    detector $j$. The sky map pixels are square but the detectors are
    quadrilaterals because of the distortion of the optics. The
    distortion is exaggerated in this sketch.}
  \label{fig:grid}
\end{figure}

The PACS data are a set of samples of the bolometers taken at
different times and thus at different positions in the sky (in scan
mode). Each sample is proportional to the sky surface brightness and
to the area of the detector projected on the sky sphere (see Figure
\ref{fig:grid}). Thus, PACS data and the flux of the sky are linearly
related and we can write equation (\ref{eq:proj}) where $\yb$ stands
for the data, $\xb$ denotes the sky pixels values and $\Pb$ is the
projector. The noise is considered additive and is noted $\nb$.
\begin{equation}
  \label{eq:proj}
  \yb = \Pb \xb + \nb
\end{equation}

We take into account the distortion of the field of view (FOV) by the
optics of the instrument in the projector. The gains of the bolometers
can be taken into account too. In our formalism, we would simply
multiply $\Pb$ by a diagonal matrix with the gain map on the diagonal
(assuming that the gain does not vary with time, the gain map would
simply be the flat-field repeated n$_f$ times, where n$_f$ is the
number of frames). The compression will be taken into account
separately.

In this paper, we used an implementation of the PACS instrument model
provided by \cite{Chanial10Tamasis} in preparation.

\section{Map-making}

The map-making stage is the inverse problem associated with equation
(\ref{eq:proj}), that is the estimation of $\xb$ knowing $\yb$. It can
be computed using classical estimation methods. We now review some of
the methods that have been implemented for the map-making stage.

\subsection{PACS pipeline}
\label{subsec:pipeline}

The Herschel Interactive Processing Environment (HIPE)\footnote{"HIPE"
  is a joint development by the Herschel Science Ground Segment
  Consortium, consisting of ESA, the NASA Herschel Science Center, and
  the HIFI, PACS and SPIRE consortia. (See
  http://herschel.esac.esa.int/DpHipeContributors.shtml).}  offers
back-projection of the data onto the sky map for the map-making
pipeline step.  In order to guarantee the conservation of the flux,
the map must be divided by the coverage map (that is the number of
times a pixel of the sky map has been seen by the detectors). The sky
map estimation using the pipeline is thus described by equation
(\ref{eq:pipeline}).
\begin{equation}
  \label{eq:pipeline}
  \hat{\xb}_{pipeline} = \frac{\Pb^T \yb}{\Pb^T \mathbf{1}}
\end{equation}

Estimated sky maps are noted with a $\hat{\cdot}$. $\cdot^T$ is the
transpose of a matrix, $\mathbf{1}$ is a vector of ones. The
division of two vectors here is the term by term division of their
coefficients.

This method is very fast and straightforward to implement. But it does
not provide a very accurate estimate of the sky. To illustrate this,
one could compare this estimate to the least-squares estimate of the
sky map. The least-squares estimate minimizes the criterion $ \|\yb -
\Pb\xb \|^2$ and can be written in closed form as in equation
(\ref{eq:ls}). The least-squares solution can be seen as the maximum
likelihood solution under the assumption of independent and
identically distributed Gaussian noise.
\begin{equation}
  \label{eq:ls}
  \hat{\xb}_{lsq} = \left( \Pb^T \Pb \right)^{-1} \Pb^T \yb
\end{equation}

The pipeline solution can now be interpreted as a very crude
approximation of the least-squares solution by replacing the inversion
of $\Pb^T \Pb $ by the inversion of the coverage map. This
approximation ignores the correlations introduced by the projection
matrix. In a more physical interpretation, one can say that the
pipeline solution does not take into account the possibility to make
use of super-resolution from the PACS data. Reducing the pixel size in
the pipeline sky map will not help on this matter. The projection
matrix inversion is required. In fact, the PSF is not always sampled
at the Nyquist-Shannon frequency and it is thus possible to obtain a
better resolution in the sky map than in the raw data. As can be seen
in Table \ref{tab:filters} this is mainly beneficial for the blue
filter (70 $\mu$m) in which aliasing occurs the most.

\begin{table}
  \centering
  \begin{tabular}{|c|c|c|c|}
    \hline
    \rule{0pt}{1.05em}filter & 70 $\mathbf{\mu}$m &
    \rule{0pt}{1.05em}100 $\mathbf{\mu}$m & 160 $\mathbf{\mu}$m \\
    \hline
    \rule{0pt}{1.05em}wavelength range ($\mu$m) & 60--85 & 85--130 & 130--210 \\
    \hline
    \rule{0pt}{1.05em}resolution ($\lambda / \Delta\lambda$)
    & \multicolumn{2}{c|}{ $\sim$ 3} & $\sim$ 2 \\
    \hline
    \rule{0pt}{1.05em}pixel size (arcsec) & \multicolumn{2}{c|}{3.2} & 6.4 \\
    \hline
    \rule{0pt}{1.05em}FOV (arcmin) & \multicolumn{3}{c|}{3.5 $\times$ 1.75} \\
    \hline
    \rule{0pt}{1.05em}FWHM & 5.2 & 7.7 & 12 \\
    \hline
  \end{tabular}
  \caption{PACS photometer overall characteristics/performances
    (taken from the PACS Observer's Manual). The resolution is
    the spectral resolution of the bandwidth.
  }
  \label{tab:filters}
\end{table}

\subsection{MADmap}

There is another approach called MADmap \citep{cantalupo2010madmap}
which tries to better estimate the sky map. It makes use of a
preconditioned conjugate gradient in order to efficiently estimate the
least-squares solution. Furthermore, MADmap adds a more accurate model
of the pink noise.  The noise is modeled as a multivariate normal
distribution with a covariance matrix $\Db$ diagonal in Fourier space
(equation (\ref{eq:cov})). Noting $\Fb$ the matrix of the Fourier
transform along the time axis, $\Fb^\dag$ its conjugate transpose (and
also its inverse), $\diag$ the operator which transforms vectors into
diagonal matrices and $\tilde \cdot$ the Fourier transform of a
vector or a matrix, we have:
\begin{equation}
  \label{eq:cov}
  \nb = \mathcal{N} (\mathbf{0}, \Db)
  \text{, } 
  \Db = \Fb^\dag \tilde{\Db} \Fb = \Fb^\dag \diag(\tilde{\db}) \Fb
\end{equation}
This allows to store the full covariance matrix in an array of the
size of the data, by storing the diagonal elements of the Fourier
transform of $\Db$, which we note here $\tilde{\db}$.

The map estimation in this case can be rewritten as a simple
least-squares estimation following equation (\ref{eq:madmap}).
\begin{eqnarray}
  \label{eq:madmap}
  \yb' & \leftarrow & \Db^{\frac{1}{2}} \yb  \nonumber \\
  \Pb' & \leftarrow & \Db^{\frac{1}{2}} \Pb \nonumber \\
  \hat{\xb}_{MADmap} & = & 
  \left(\Pb^T \Db \Pb \right)^{-1} \Pb^T \Db^{\frac{1}{2}} \yb 
  \nonumber \\
  & = & \left(\Pb'^T \Pb' \right)^{-1} \Pb'^T \yb'
\end{eqnarray}
$\Db^\frac{1}{2}$ is such that $\Db^{\frac{1}{2}} \Db^{\frac{1}{2}} = \Db$.

The difficulty resides in finding a proper estimation for the noise
covariance matrix. One has to assume that the noise covariance matrix
is time invariant at very long time scales in order to have enough
data samples for its estimation. In addition, one has to assume that
there is only noise in the data taken to estimate the noise covariance
matrix. 

For Herschel/PACS data, we have data sets close enough to pure noise
data so that estimating the noise statistics is not an issue.  In
principle, the noise covariance matrix should be of the size of the
data set since the $1 / f$ noise extends to the largest scales. In
practice we can perform a median filtering at very large scales and
use the noise covariance matrix estimation for intermediate
scales. The median filter window can be made large enough to avoid
filtering any scale corresponding to the physical sources of
interest. This is easier for unresolved point sources but even for
extended emission the scales of interest do not exceed the size of the
map.

Furthermore, the current MADmap implementation does not make use of an
accurate projection matrix but assumes a one-to-one pixel association
between sky map pixel and data pixel.  This limits the capacity of
MADmap to take advantage of the super-resolution possibilities of PACS
data since the projection matrix is not accurate. Another limitation
of MADmap is due to the temporal-only Fourier transform. Because of
it, the covariance matrix can only model temporal correlation,
neglecting detector to detector correlations, which have to be removed
by some pre-processing. However, most are gone after median filtering.

\section{Compressed-Sensing}
\label{sec:cs}

\subsection{Theory}
\label{subsec:theory}

It is now well known that natural images and astronomical images can
be sparsely represented in wavelet bases. This finding has been
successfully applied to a number of problems and to compression in
particular (JPEG2000 \citep{skodras2001jpeg} or Dirac video
compression \citep{borer2005dirac}). But, until recently, it had not
been realized that it is possible to make use of this knowledge to
derive more efficient data acquisition schemes. This is precisely the
idea behind the compressed-sensing (CS) theory. Compressed-sensing
states that if a signal is sparse in a given basis, it can be
recovered using fewer samples than would be required by the
Shannon-Nyquist theorem, hence the term compressed-sensing. See
\citet{candes2008people} for a tutorial introducing compressed-sensing
theory or \citet{candès2006compressive} and
\citet{donoho2006compressed} for seminal articles. You can also look
at \citet{starck:book10} which have a lot of example applications in
astrophysics.

More formally, let us assume that our signal of interest $\xb$ is
sparse in the basis $\Bb$ and we are taking samples $\yb$ through the
measuring matrix $\Ab$ (i.e. $\yb = \Ab \xb$). In this context sparse
means that $\alphab = \Bb\xb$ have few non-zero elements. $\alphab$ is
said to be approximately sparse if it has few ``significant'' elements
(far from zero).  The theory proposes two categories of measurement
matrices that allow precise estimation of $\xb$ : incoherent matrices
and random matrices. Random measurements are likely to give incoherent
measurements. Incoherence is the fundamental requirement for $\Ab$ in
the compressed-sensing framework.  Quantitatively, it means that the
mutual coherence $\mu = max_{j,k} \|\mb_j^T \bb_k \|$ is low where
$\mb_j$ and $\bb_k$ are the columns of $\Ab$ and $\Bb$ respectively.
Under the assumptions of exact sparsity and incoherence, it can be
shown that $\xb$ can be estimated exactly using few
measurements. Precisely, if $m$ is the number of measurements, s is
the number of non-zero elements and n the number of unknowns, we need
$m \ge C \cdot \mu^2 \cdot s \cdot log(n)$, where $C$ is a constant
\citep{candes2008introduction}.  Under the approximate sparsity
hypothesis, good performances can be demonstrated too
\citep{stojnic2008compressed}.

In order to get the benefits of CS it is assumed that it is possible
to solve the inverse problem under the assumption of sparsity (i.e. to
find a sparse solution to the inverse problem). In other word, we need
to find the solution of problems of the type of equation
(\ref{eq:sparse}).
\begin{equation}
  \label{eq:sparse}
  \hat{\xb}_{CS} = \Argmin{\xb}{ \|\Bb\xb \|_0} 
  \text{ subject to }  \|\yb - \Ab\xb  \|^2  < \epsilon
\end{equation}
where $\| \cdot \|_0$ is the L$_0$ norm which is the number of
non-zero elements of the vector and $\epsilon$ is a small value linked
to the variance of the signal. This estimation problem is of
combinatorial complexity (all solutions need to be tested) and thus is
intractable for our applications.

Since equation (\ref{eq:sparse}) is of combinatorial complexity, it is
often replaced by equation (\ref{eq:sparse_l1}) (see
\citet{bruckstein2009sparse}), where the L$_0$ norm has been replaced
by the L$_1$ norm ( $\| \|_1$ is the L$_1$ norm), or the sum of the
absolute values of the coefficients.
\begin{equation}
  \label{eq:sparse_l1}
  \hat{\xb}_{CS} = \Argmin{\xb}{ \|\Bb\xb \|_1}
  \text{ subject to }  \|\yb - \Ab\xb  \|^2  < \epsilon
\end{equation}

We can choose to write the estimation problem as in equation
(\ref{eq:sparse_approx}). It can be seen as a reformulation of
equation (\ref{eq:sparse_l1}) using Lagrange multipliers.
Statistically, it can also be seen as the co-log-posterior (the
opposite of the logarithm of the posterior distribution) of an inverse
problem assuming a normal likelihood and a Laplace prior.
\begin{equation}
  \label{eq:sparse_approx}
  \hat{\xb}_{CS} = \Argmin{\xb}{ \|\yb - \Ab\xb  \|^2 
    + \lambda \|\Bb\xb \|_1}
\end{equation}

Note that this modeling of the problem ignores some artifacts of PACS
data such as glitches and pink noise. We will see how we can handle it
in preprocessing.

\subsection{PACS implementation of compressed-sensing}
\label{subsec:pacs_cs}

We now describe how we implement the CS theory to PACS data in
practice. Let us stress that we cannot modify the way Herschel
observes and that we are severely limited by computer resources on
board. We choose to apply a linear compression matrix $\Cb$ to each
frame. $\Cb$ is the compressed-sensing part of the acquisition model.
The new inverse problem can now be described as in equation
(\ref{eq:cs_model}).
\begin{equation}
  \label{eq:cs_model}
  \zb = \Cb\Pb\xb + \nb
\end{equation}

\begin{figure}
  \centering
  \includegraphics[width=\linewidth]{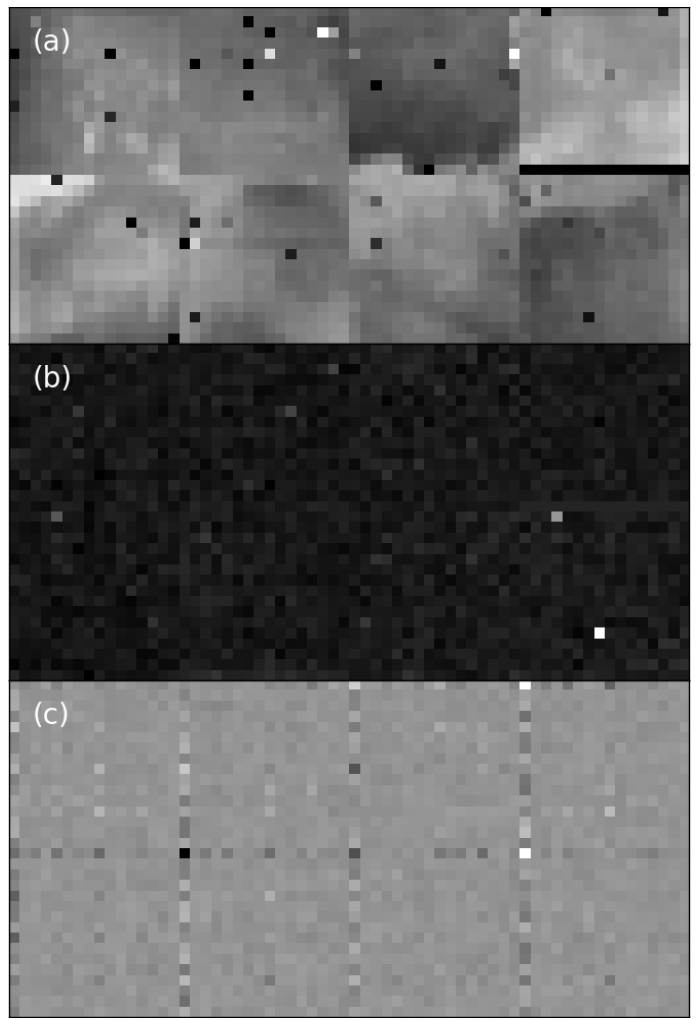}
  \caption{(a) Frame of the PACS imaging camera in the blue band.
    The darker and brighter isolated pixels are bad pixels.  There is
    one complete bad line (at the bottom of the last matrix, ordered
    from the left to the right and from the bottom to the top). The
    observable structure is mainly the offset. Discontinuities in the
    offset correspond to the borders of the matrices of bolometers.
    Note that there are physical gaps between matrices as well as
    rotation and distortion.  The matrix used in the transparent mode
    is the sixth one.  (b) The same frame with the offset removed. The
    brighter isolated pixels are glitches.  (c) The same frame after
    Hadamard transform. The offsets have not been removed. }
  \label{fig:data}
\end{figure}

Due to severe on-board computational constraints we cannot make use of
random measurement matrices nor wavelet transforms. We choose the
Hadamard transform which can be implemented in linear complexity using
the fast Hadamard transform algorithm \citep{pratt1969hadamard}. The
compression is then made using pseudo-random decimation of the
Hadamard transform of the frame. The Hadamard transform has been
chosen since it mixes uniformly all the pixels. Thus, in conjunction
with the decimation, it is a good approximation of a random matrix. To
illustrate this, one can look at the Figure \ref{fig:data} which
displays the result of the Hadamard transform of a single frame of
PACS (without removing the offset).

Note that our method is not CS in a strict meaning. Because we have
more data than unknowns, we cannot tell from the CS theory which
compression matrix is optimal. But we think CS friendly compression
matrices are worth being investigated as there is no other theory that
point to other optimal compression matrices under sparsity
assumptions. We plan to test a wider range of compression matrices in
future works.

We also point out that the standard compression scheme of PACS scans
(averaging) can also be modeled as in equation (\ref{eq:cs_model})
replacing $\Cb$ by an averaging matrix.  The exact same inversion
scheme can be applied to both the CS inversion problem and the
standard inversion problem allowing a clear comparison by separating
effects due to the acquisition matrix from those due to the sparsity
constraint.

Equation (\ref{eq:averaging_matrix}) shows the averaging compression
that is usually performed on PACS data. In this equation $\Ib$ is the
identity matrix of size the number of detectors.

\begin{equation}
  \Cb_{AV} = \left(
  \begin{tabular}{ccccccc}
    \Ib & \dots &\Ib &&&&\\
     &&& $\ddots$ &&&\\
    &&&&\Ib & \dots &\Ib \\
  \end{tabular}
  \right)
  \label{eq:averaging_matrix}
\end{equation}

Equation (\ref{eq:cs_matrix}) shows the compressed-sensing compression
that we implemented. $\Hb$ is the Hadamard transform of one frame, and
$\Mb_i$ is a decimation matrix. A decimation matrix is a rectangular
matrix with only ones and zeros values and exactly one non-zero
element per row. It selects the elements of the Hadamard transform to
be transmitted. For example, in the parallel mode, only one-eighth of
the values are transmitted. The decimation matrices vary from one
frame to the next to minimize the redundancy of the transmitted
information. $F$ is the compression factor.

\begin{equation}
  \small
  \Cb_{CS} = \left(
  \begin{tabular}{ccccccc}
    $\Mb_1 \Hb$ &&&&&&\\
    & $\ddots$ &&&&&\\
    && $\Mb_F \Hb$ &&&&\\
    &&& $\Mb_1 \Hb$ &&&\\
    &&&& $\ddots$ &&\\
     &&&&& $\Mb_F \Hb$  &\\
  \end{tabular}
  \right)
  \label{eq:cs_matrix}
\end{equation}

Note that all the matrices in this paper are not implemented as
matrices but rather as linear functions as it would not be feasible to
store such large matrices and prohibitive to perform matrix
multiplications.

We define sparse priors in both wavelet space and finite difference.
The criterion can then be rewritten as in equation (\ref{eq:map}),
where $\Wb$ is a wavelet transform and $\Db_x$ and $\Db_y$ are finite
difference operators along both axes of the sky map. The finite
difference operators simply compute the difference between two
adjacent pixels along one axis or the other. Thus, as a prior they
favor smooth maps. Since we minimize a $L_1$ norm, it allows for
some bigger differences, hence the map can be piecewise smooth.
\begin{eqnarray}
  \label{eq:map}
  \hat{\xb}_{CSH} &=& \Argmin{\xb}{\| \zb - \Cb\Pb\xb\|^2} \nonumber \\
    && {} + \lambda_W \|\Wb\xb \|_1
    + \lambda_x \|\Db_x\xb \|_1
    + \lambda_y \|\Db_y\xb \|_1
\end{eqnarray}

We estimate $\hat{\xb}_{CSH}$ (CSH meaning Compressed-Sensing for
Herschel) using a conjugate gradient (CG) with Huber norm. The Huber
norm is defined according to equation (\ref{eq:huber}).
\begin{equation}
  \label{eq:huber}
  \left\{
  \begin{array}{cccc}
    \mathcal{H}(x) &=& x^2 &, 
    \forall x \in \left[ - \delta, \delta \right] \\
    \mathcal{H}(x) &=& 2 \left| \delta x \right| - \delta^2 &, 
    \forall x \notin \left[ - \delta, \delta \right] \\
  \end{array}
  \right.
\end{equation}
The Huber norm is a way to approximate a L$_1$ norm ($\| \cdot \|_1$)
with a function everywhere differentiable, which is mandatory in order
to use CG methods. For this purpose, the Huber parameter $\delta$ can
be chosen very small.

Finally, we minimize equation (\ref{eq:min_csh}).
\begin{eqnarray}
  \label{eq:min_csh}
  \hat{\xb}_{CSH} &=& \Argmin{\xb}{\| \zb - \Cb\Pb\xb\|^2} \nonumber \\
   && + \lambda_W \mathcal{H}( \Wb\xb )
    + \lambda_x \mathcal{H}( \Db_x\xb )
    + \lambda_y \mathcal{H}( \Db_y\xb )
\end{eqnarray}

This optimization scheme is very similar to \citet{lustig2007sparse}
in the idea of applying CG iterations and replacing the L$_1$ norm by
a close function everywhere differentiable. 

\section{Preprocessing}
\label{sec:preprocessing}

It is important to note that despite the significant alteration of the
aspect of the data after CS compression (or any kind of linear
compression), it is still possible to apply the same preprocessing
steps that are performed in the standard PACS pipeline. Let us
demonstrate this in details.

The following is usually applied to the data before the inversion step
(back-projection or MADmap) :
\begin{itemize}
\item Removal of the offset by removing the temporal mean along each
  timeline
\item Deglitching
\item Pink noise filtering using median filtering (removes only the
  slowly drifting part in the case of MADmap)
\end{itemize}

\subsection{Offset and pink noise removal}
\label{subsec:pink_noise}

The removal of the offset is simple. Since the compression is linear,
and applied independently on each frame, one can remove the offset by
removing the temporal mean as in the averaging mode. The same applies
for the pink noise removal.

In order to mitigate the effect of pink noise in the estimated maps, a
filtering is performed during the preprocessing. Following the method
of the pipeline, we use an high-pass median filtering. This can have a
side effect of removing signal of interest at larger spatial scales in
the map. This is of little importance in fields containing mostly
point sources but can be dramatic in fields with structures at every
scale such as fields taken in the galactic plane. One of the interests
of MADmap is to be able to remove the pink noise artifacts in the maps
without affecting the large scale structures. But even with
MADmap-like methods, one still have to remove the very low frequency
component of the pink noise not taken into account by the estimated
noise covariance matrix. That is why it is important to ensure that
one can perform median filtering in a CS framework.

We perform high-pass median filtering on the CS data exactly in the
same way as it is done with the averaged data. Thanks to the linearity
of the CS compression model, the pink noise in the raw data remains
pink noise in the CS data. Thus, the median filtering affects the
noise in the CS data the same way it does in the averaged data.

The only important parameter in the median filtering is the length of
the sliding window used to compute the median in the neighborhood of
each point. If the window is too large, there will be too much pink
noise remaining in the filtered data. If it is too small, large scale
structures in the map will start to get affected. As a compromise, we
choose the window to be of the size of a scan leg, which is
approximately the largest scale in the map. In the NGC6946 maps this
corresponds very roughly to a window of 1000 samples at a sample rate
of 40 Hz at a speed of 60 arcsec/s. For the compressed data, we
divide this length by the compression factor.

\subsection{Deglitching}
\label{subsec:deglitching}

The deglitching is more complicated since it makes use of spatial
information. In order to apply deglitching in the CS case, we first
decompress the data without projecting onto the sky map (equation
(\ref{eq:decompression})).

\begin{equation}
  \label{eq:decompression}
  \hat{\yb} = (\Cb^T \Cb)^{-1} \Cb^T \zb
\end{equation}

We can then perform deglitching on $\hat{\yb}$ as we would have done
on averaged data. We perform deglitching on $\hat{\yb}$ using a method
called \textit{second level deglitching}. It consists of regrouping
all data samples that correspond to a given pixel map and masking out
the ones that are beyond a given threshold using the median absolute
deviation (MAD) method.

Note that any operation that is performed on standard data could be
applied on CS data using equation (\ref{eq:decompression}). This is a
very important advantage of using a linear compression method. Other
compression methods, such as JPEG for instance, do not satisfy this
property, since a thresholding is performed. It is also the linearity
of the compression which allows to jointly perform the decompression
and the map-making steps of the inversion and thus to obtain better
maps.

When $\Cb$ is an averaging matrix (the standard compression), the
samples that are found to be glitches are simply masked. In our case,
we define a masking matrix $\Mb$ which is simply an identity matrix
with zeros on diagonal elements corresponding to where values are
masked. We can finally perform the inversion replacing $\Cb\Pb$ by
$\Cb\Mb\Pb$ in equation (\ref{eq:min_csh}). This way, the inversion is
done without altering the data set but entries corresponding to
glitches are not taken into account in the inversion. 

As can be seen in the maps in Figure \ref{fig:maps}, this is an
effective method since no glitch remains visible. To ensure optimal
detectability of the glitches it is best to perform pink noise removal
beforehand. One can use a shorter window for the median filtering to
better remove the pink noise only for the deglitching step. Indeed, in
this case, it does not matter if the map structures are affected since
the filtered data are only used to compute a mask for the
glitches. One can then apply the mask on the data filtered with a
larger window to preserve large scale structures.

\subsection{Conclusion}

To summarize, we can perform the same processing as the pipeline in
the CS mode. There is no obstacle remaining which would prevent us
from applying CS to PACS data. Let us see now the results we obtained
with CS on real data.

\section{Results}
\label{sec:results}

We present results of sky map estimation using PACS data observations
in transparent mode. The different modes of compression are
simulated. (PL) is the pipeline mode, i.e. a simple weighted
backprojection ignoring that the data have been compressed (see
equation \ref{eq:pipeline}). Compression is done by averaging. (PLI)
is a full inversion of the pipeline model. It means it does not take
into account the correct compression model. Data for (PLI) have been
compressed by averaging. In other words (PLI) is the same as (PL)
replacing the backprojection with a full inversion. (ACI) is the
inversion of the model including the compression by averaging, (HCI)
is the inversion with Hadamard compression and (NOC) is the inversion
of the full data set (not compressed). All these methods but (PL) use
the same sparse inversion algorithm described in section
\ref{subsec:pacs_cs}.

\begin{figure*}
  \centering
  \includegraphics[width=.9\linewidth]{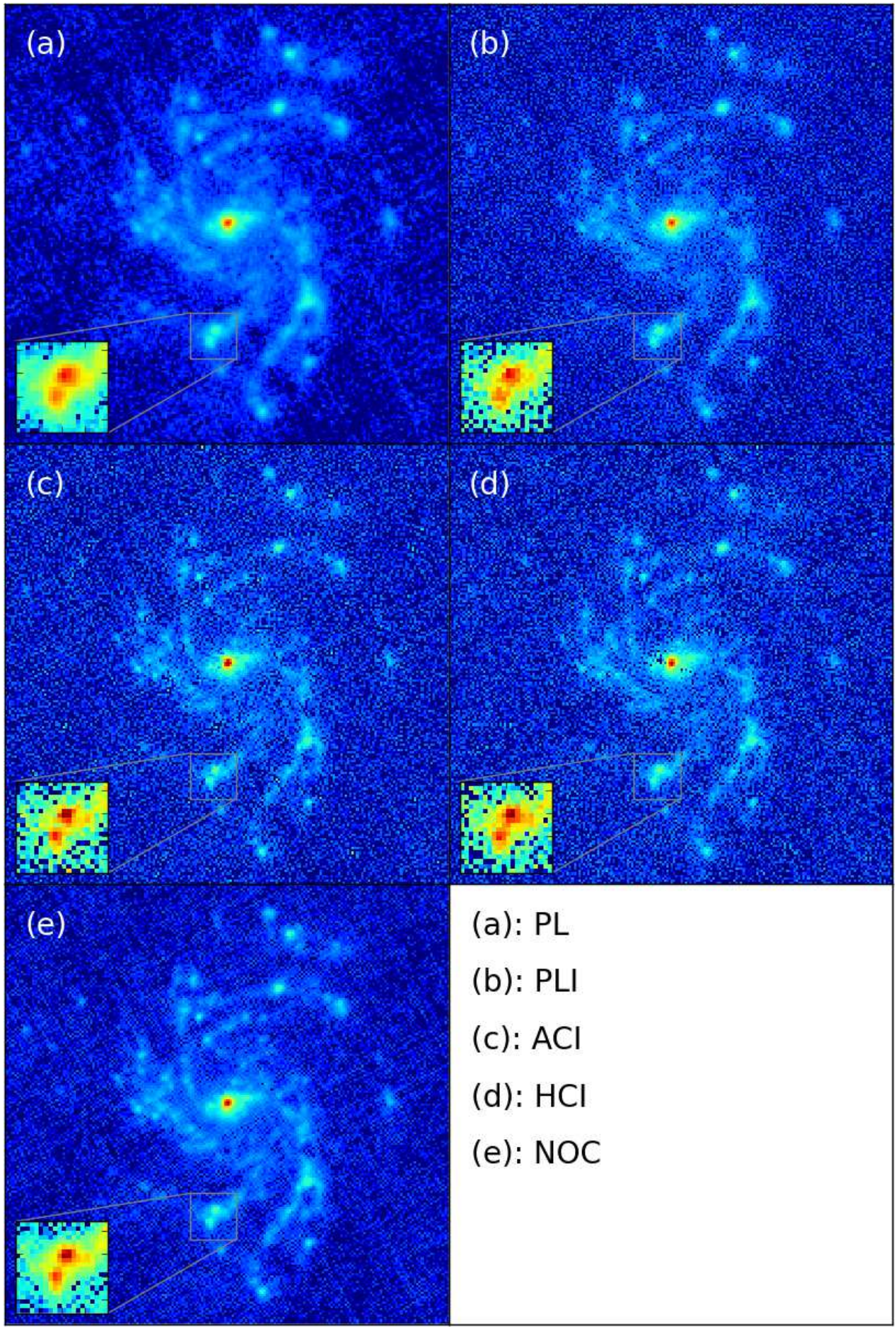}
  \caption{ Map estimations of NGC 6946 using (a) the pipeline (PL)
    back-projection method , (b) the pipeline model inversion (PLI)
    (without taking into account the compression), (c) averaging
    compression inversion (ACI), (d) Hadamard compression inversion
    (HCI) and (e) reference map without any compression (NOC).  Maps
    are presented on a fourth root intensity scale. All maps have the
    same scale. }
  \label{fig:maps}
\end{figure*}

\begin{figure}
  \centering
  \includegraphics[width=\linewidth]{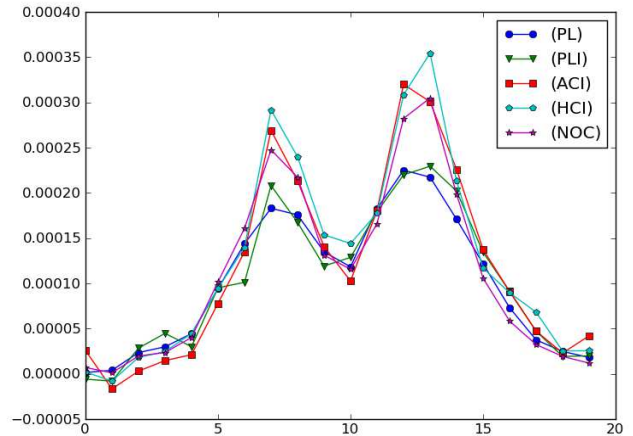}
  \caption{Line cuts along the two sources displayed in the zoom of
    figure \ref{fig:maps}. Cuts are made by two-dimensional linear
    interpolation of the maps. The x axis is the pixel index of the
    interpolation.  The y axis is the intensity in Jy. 
  }
  \label{fig:line_cuts}
\end{figure}

The compression factor on the acquisition side is always 8 in the
following results except for the reference map (Figure
\ref{fig:maps}e) for which no compression is applied. The data have
been taken in fast scan mode (60 arcsec/s). Those are the conditions
that give the most important loss of resolution in standard
(averaging) mode since the blurring of each frame is the most
important. The data set is made of two cross scans in the blue band of
approximately 60000 frames each. Each frame has 256 pixels (instead of
2048 because of the transparent mode). The estimated map is made of
$192 \times 192$ pixels of 3 arcsec each, resulting in a 10 arcmin$^2$
map. The object observed is the galaxy NGC6946 from the KINGFISH
program (P.I. R. Kennicutt) but the observations are dedicated
calibration observations. Results are presented in Figure
\ref{fig:maps}. Cuts along a line going through the two sources in the
zoom are presented in Figure \ref{fig:line_cuts}.

To illustrate the impact on the spatial resolution of the different
reconstruction methods, we list in Table \ref{tab:gaussfit} the mean
of the full width at half maximum (FWHM, in pixels) derived from
Gaussian fits to four compact sources selected on the field. Although
these sources are all not-strictly point-like, comparison of the
achieved mean FWHM is quite instructive.

\begin{table}
  \centering
  \small
  \begin{tabular}{|c|c|c|c|c|c|}
    \hline
    FWHM (pixels) & (PL) & (PLI) & (ACI) & (HCI) & (NOC) \\
    \hline
    mean of 4 sources &  4.30 & 4.01 & 2.75 & 3.27 & 3.15 \\
    \hline
  \end{tabular}
  \caption{FWHM of sources in pixels in Figure \ref{fig:maps}
    estimated using 2-dimensional Gaussian fits of 4 isolated sources
    randomly chosen from the map.}
  \label{tab:gaussfit}
\end{table}

To compare quantitatively the global results, we made radial profiles of
NGC 6946 using maps from each method. Radial profiles were obtained
using galaxy parameters from the NASA Extragalactic Database (NED) and
summed up in Table \ref{tab:ngc6946}. Surface brightness is averaged
over pixels contained in concentric ellipses. Background values are
estimated using corners of the maps where no source can be seen. The
standard deviation of the pixels in the background areas give an
estimate of the standard deviation of the noise in the whole
map. Those values are reported in Table \ref{tab:background}.

\begin{table}
  \centering
  \footnotesize
  \begin{tabular}{|c|c|c|c|c|c|}
    \hline
    Maps & (PL) & (PLI) & (ACI) & (HCI) & (NOC) \\
    \hline
    std (10$^{-6}$ Jy) & 4.84 & 14.3& 19.2 & 20.3 & 8.3 \\
    \hline
  \end{tabular}
  \caption{Standard deviation for each map in Figure
    \ref{fig:maps}. Those values are estimated from the corners of the
    maps ($\sim$ 50 $\times$ 50 pixels area at each corner for a total
    of $\sim$ 10000 pixels in each map).  }
  \label{tab:background}
\end{table}

\begin{figure*}
  \centering
  \includegraphics[width=.9\linewidth]{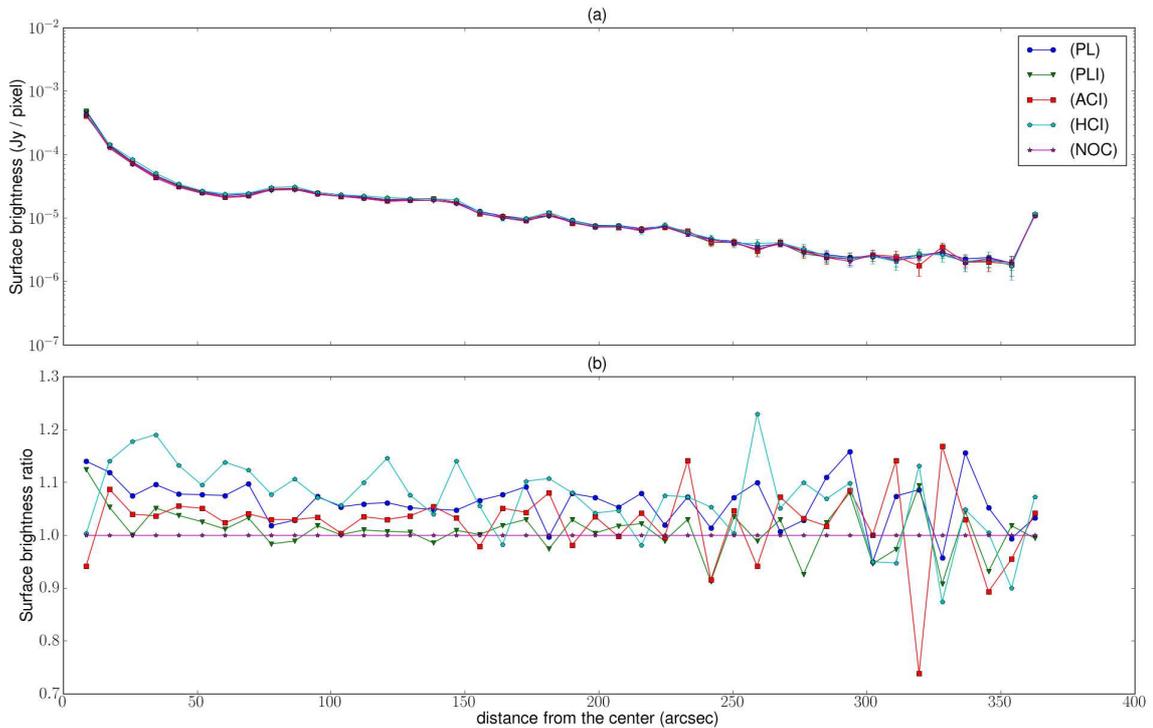}
  \caption{(a) Radial profiles of NGC 6946 in surface brightness (Jy)
    as a function of distance from the center of the galaxy (arcsec).
    Error bars are inferred from the background measurements reported
    in Table \ref{tab:background} and from the number of pixels at in
    the ellipses at each radius.  (b) Ratio of radial profiles of the
    different maps over the radial profile of the reference map
    obtained without compression.  Radial profiles are obtained from
    the maps in Figure \ref{fig:maps}. Acronyms refers to Figure
    \ref{fig:maps}.}
  \label{fig:profiles}
\end{figure*}

Results are presented in Figure \ref{fig:profiles}. Figure
\ref{fig:profiles}a shows radial profiles corresponding to each map
from figure \ref{fig:maps}. Those profiles are supplemented with
error bars estimated from the background measurements given in Table
\ref{tab:background}. We assumed normal noise meaning that the
standard deviation of the surface brightness at each radius is given
by $\sigma_{profile} = \frac{\sigma_{bkg}}{\sqrt{N(r)}}$, where $N(r)$
is the number of pixels in at radius $r$ and $\sigma_{bkg}$ is the
standard deviation of the pixels in the background.

Figure \ref{fig:profiles}b shows the ratio of those radial profiles
over the radial profile of the reference map without compression. Note
that the reference map is not guaranteed to be the ground truth. In
particular, it is affected with the potential issues of the median
filtering (used to remove pink noise) as much as the other
maps. However, we assume it is closer to the ground truth since it
does not suffer from a loss of information due to compression.

We deduced the total flux of NGC6946 from the maps in Figure
\ref{fig:maps} and we display the result in Table \ref{tab:total_flux}.

\begin{table}
  \centering
  \begin{tabular}{|c|c|c|c|}
    \hline
    PA (\degree) & i (\degree)
    & ra (\degree) & dec (\degree) \\
    \hline
    52.5 & 46 & 308.71 & 60.153 \\
    \hline
  \end{tabular}
  \caption{Parameters of the NGC 6946 galaxy. PA is the position
    angle, i is the inclination, ra is the right ascension and dec is
    the declination.  }
  \label{tab:ngc6946}
\end{table}

\begin{table}
  \centering
  \begin{tabular}{|c|c|c|c|c|c|}
    \hline
    Maps & (PL) & (PLI) & (ACI) & (HCI) & (NOC) \\
    \hline
    flux (Jy) & 0.65 & 0.62 & 0.64 & 0.67 & 0.63 \\
    \hline
  \end{tabular}
  \caption{Total flux of NGC 6946 in Jy as estimated from the maps
    of Figure \ref{fig:maps}. }
  \label{tab:total_flux}
\end{table}

\section{Discussion}
\label{sec:discussion}

Comparing Figures \ref{fig:maps}a and \ref{fig:maps}b, there is a
small gain in resolution. The FWHM estimate in table
\ref{tab:gaussfit} illustrates this small gain since the FWHM drops
from 4.3 pixels to 4.0 pixels. This gain comes from the inversion of
the projection matrix. Physically, it means that we used the pointing
information to get a super-resolved map. Note that this results holds
only for the blue band which is not sampled at the Nyquist frequency.

Looking at Figure \ref{fig:maps}, maps ACI and HCI are better resolved
than maps PL and PLI. This is supported by the quantitative results
presented in Table \ref{tab:gaussfit}. Figure \ref{fig:line_cuts}
shows that this gain in resolution allows close sources to be better
separated. From Table \ref{tab:gaussfit}, the gain in resolution is
approximately 1 pixel, or 3 arcseconds. This gain is due to the
inversion of linear models which take into account the compression
step and to the inclusion of sparsity priors in the inversion
algorithm. This is common to both ACI and HCI and not intrinsic to CS.

Comparing ACI and HCI in Figure \ref{fig:maps}, it appears that the
structure of the remaining noise is different. HCI does not exhibit
linear patterns in the noise, following the scan direction.  This is
due to the mixing of the pink noise component of each detector done by
the Hadamard compression. Furthermore, Table \ref{tab:gaussfit} shows
that ACI has a mean FWHM below NOC. This is unexpected since NOC has
more data than ACI. But this can be explained by the median filtering
performed to filter the pink noise. Median filtering can have a small
effect on compact sources even when large sliding windows are used. On
the opposite, HCI seems more robust to this median filtering since the
mean FWHM in the HCI map is just slightly above the NOC mean FWHM and
not below it.

Of course, these lossy compression schemes still imply a loss of
information but it mostly translates into an increase of the
uncertainty of the map estimation (i.e. the noise). This is true for
both compression matrices.

Note, however, that comparing results between CS and averaging
compression matrices does not allow to conclude on the absolute
performance of compressed-sensing. To conclude on this point, a study
is needed to find the best CS compression matrix. In this paper we
only try to highlight the applicability of CS methods to real data
from a space observatory.

The amount of residual noise can seem surprising if one does not
remember that the noise in PACS data is mostly a pink noise, and thus
is not well taken into account in our inversion model. We need to make
use of the noise covariance matrix in our inversion scheme.  In order
to do so, we can update equation (\ref{eq:min_csh}) by taking into
account a noise covariance matrix. This can be done using methods
inspired by MADmap. We will investigate this possibility in a
forthcoming article. However, since the resolution is increased in our
results, the noise increase cannot be directly translated into a
sensitivity decrease.

In figure (\ref{fig:profiles}a), we can see that the radial profiles
obtained using the maps displayed in figure (\ref{fig:maps}) are very
close to each other. It gives confidence in the capability of our
methods to preserve the flux in the maps at larger scales. In
particular, this shows that the high-pass filtering is not affecting
the profiles differently for CS than for averaging compression. It
validates the way we perform this preprocessing step in the CS
pipeline.

Figure (\ref{fig:profiles}b) shows the relative discrepancies between
the profiles. They are higher further from the center since the
surface brightness are very low, but they are consistent with the
uncertainties. Again the CS compression scheme and the averaging
compression scheme are very similar.

Profiles from compressed maps seem to be systematically slightly
over-estimated between 0 and 150 arcseconds from the center,
especially for (HCI) and (PL) maps. However, as shown in Table
\ref{tab:total_flux} the total flux of the galaxy computed for each
map are within 7 \% of the flux computed from the reference map. This
result is in full accordance with the photometric accuracy quoted by
the PACS team \citep{poglitsch2010photodetector}.

We presented results for a resolved galaxy only but our method could
benefit other objects. For example, deep survey observations are
limited by the confusion limit at which the presence of numerous
unresolved sources appear as noise. We plan to see if the confusion
limit could be improved with our approach in future work. The gain in
resolution could also help with regard to the source blending issue
and could help improve the precision of catalogs of isolated
unresolved sources (as estimated by SExtractor
\citep{bertin1996sextractor}, for instance). This could be tested on
simulated field of sources with known positions. See also
\cite{Pence2010Optimal} for a study on compression method using this
idea.

\section{Conclusions}
\label{sec:conclusions}

In this paper we showed that methods inspired by the
compressed-sensing theory can successfully be applied to real
Herschel/PACS data, taking into account all the instrumental
effects. For this purpose, special PACS data was acquired in the
``transparent mode'' in which no compression is applied. We then
performed compressed-sensing compression scheme as well as the
standard averaging compression scheme on the very same data which
allows for an unbiased comparison. The data are affected by various
artifacts such as pink noise and glitches. Handling of these artifacts
is well known on standard averaged data but we show that the same kind
of techniques can successfully be applied on compressed-sensing data.

Furthermore, we showed that it is possible to significantly improve
the resolution in the sky maps compared to what can be obtained using
the pipeline. This is done by taking into account the compression
matrix in the acquisition model and by performing a proper inversion
of the corresponding linear problem under sparsity constraints.

For now the resulting sky maps in CS and averaging compression modes
are of the same quality. However, it is important to note that we may
not be applying the best CS method. In particular, the Hadamard
transform is only one compression matrix among others which satisfies
the requirements for a CS application. Thus, this article does not in
any way conclude on the performance of CS over standard
compression. We can only conclude that this particular approach
inspired by the CS framework and this particular Hadamard transform
give as good results as the averaging compression matrix. But choosing
another compression matrix could further improve the CS results over
averaging.

We plan to more systematically analyse other choices for the
compression matrix in a further article. Other improvements can be
expected by performing deconvolution of the instrument PSF during the
map-making stage and including the noise covariance matrix as in
MADmap.

\section{Acknowledgements}
\label{sec:acknowledgements}

We thank the ASTRONET consortium for funding this work through the CSH
and TAMASIS projects. Roland Ottensamer acknowledges funding by the
Austrian Science Fund FWF under project number I163-N16.

\bibliographystyle{dep/aa}
\bibliography{BarbeyCSH}

\end{document}